\documentclass[showpacs,showkeys,twoside,eqsecnum,twocolumn]{revtex4} 
\usepackage{amsfonts,amsmath,amssymb,bm,hyperref,graphicx}% for the archive

\begin{document}

%\preprint{E-print archive: hep-ph/0701209}

\title{Evolution of Mixed Dirac Particles Interacting \\
with an External Magnetic Field}

\author{Maxim Dvornikov$^{a,b}$}
\email{dvmaxim@cc.jyu.fi}
\author{Jukka Maalampi$^{a,c}$}
\email{maalampi@cc.jyu.fi}
\affiliation{$^a$Department of Physics,
P.O.~Box~35, FIN-40014 University of Jyv\"askyl\"a, Finland;\\
$^b$IZMIRAN, 142190, Troitsk, Moscow region, Russia;\\
$^c$Helsinki Institute of Physics, P.O.~Box~64, FIN-00014
University of Helsinki, Finland}

%
%\email{maxdvo@izmiran.ru}

\date{\today}

\begin{abstract}
We study in the framework of relativistic quantum mechanics the
evolution of a system of two Dirac neutrinos that mix with each
other and have non-vanishing magnetic moments. The dynamics of
this system in an external magnetic field is determined by solving
the Pauli-Dirac equation with a given initial condition. We
consider first neutrino spin-flavor oscillations in a constant
magnetic field and derive an analytical expression for the
transition probability of spin-flavor conversion in the limit of
small magnetic interactions. We then investigate ultrarelativistic
neutrinos in an transversal magnetic field and derive their wave
functions and transition probabilities with  no limitation for the
size of transition magnetic moments.  Although we  consider
neutrinos, our formalism is straightforwardly applicable to any
spin-$1/2$ particles.
\end{abstract}



\maketitle

\section{Introduction}

In the neutrino oscillation phenomena observed so far oscillations
occur between different neutrino flavors $\nu_{\lambda}$
($\lambda=e,\mu,\tau$) so that an active left-handed neutrino goes
to another active left-handed neutrino (e.g.
$\nu_{e}^\mathrm{L}\leftrightarrow \nu_{\mu}^\mathrm{L}$). In some
situations another type of oscillations is possible, namely
spin-oscillation, where oscillation happens between an active
left-handed neutrino and its inert right-handed counterpart (e.g.
$\nu_{e}^\mathrm{L}\leftrightarrow \nu_{e}^\mathrm{R}$). Also a
combination of these two oscillation types, so called  spin-flavor
oscillations, can happen.  There oscillations take place between
an active left-handed neutrino and an inert right-handed neutrino
of different flavor (e.g. $\nu_{e}^\mathrm{L}\leftrightarrow
\nu_{\mu}^\mathrm{R}$)~\cite{qmSFO,LimMar88}.

In this paper we shall consider spin-flavor oscillations in an external magnetic %%@
field. We assume that neutrinos are Dirac particles with non-vanishing magnetic %%@
moments. Let us remind that Dirac particles can have both ordinary (diagonal) %%@
magnetic moments and transition (non-diagonal) magnetic moments, whereas for %%@
Majorana particles only transition magnetic moments are possible~\cite{Fukugita}. %%@
Note that despite the recent claims of the experimental discovery of the Majorana %%@
nature of neutrinos~\cite{KleKriDieChk04}, it is still an open question whether %%@
neutrinos are Dirac or Majorana particles~\cite{EllEng04}. We shall use the %%@
formalism of relativistic quantum mechanics, that is, we use the Dirac equation as %%@
a starting point, which is a proper approach for spin-$1/2$ particles. One of us %%@
(MD) has previously used this formalism for studying neutrino flavor oscillations %%@
in vacuum~\cite{FOvac} and in an external axial-vector field~\cite{Dvo06EPJC} %%@
(neutrino interaction with matter). Neutrino spin-flavor oscillations in %%@
electromagnetic fields of various configurations have been studied in %%@
Refs.~\cite{SFOwe}. The present paper is a continuation of these previous works.

Let us consider a system of two Dirac neutrinos with non-vanishing masses and %%@
magnetic moments. In general, both the mass matrix and the matrix of magnetic %%@
moments  are non-diagonal in the flavor basis of neutrino wave functions. When the %%@
mass part of the Hamiltonian is diagonalized by a unitary transformation and the %%@
original flavor basis is replaced by a new set of wave functions in the mass %%@
eigenstate basis, the matrix of magnetic moments is generally not diagonal in the %%@
new basis. This means that there will be transition magnetic moments between the %%@
mass eigenstates. We will consider magnetic moment matrices of neutrinos in various %%@
bases in Sec.~\ref{EDnu}. In Sec.~\ref{PT} we will discuss the situation where the %%@
resulting  magnetic moment matrix is close to diagonal, i.e. the transition %%@
magnetic moments are small compared with the diagonal magnetic moments  in the mass %%@
eigenstate basis, in which case one can apply the formalism developed in %%@
Refs.~\cite{FOvac,Dvo06EPJC}.  In Sec.~\ref{BPT} we will consider the effects of %%@
transition magnetic moments for ultrarelativistic neutrinos  in a transversal %%@
magnetic field with no limitations on the size of any magnetic moments. In %%@
particular, we apply our result for studying the situation, where the transition %%@
magnetic moment is large compared with the diagonal ones.  We will derive the %%@
transition probability for the process like %%@
$\nu_{\beta}^\mathrm{L}\to\nu_{\alpha}^\mathrm{R}$, where $\alpha,\beta$ denote two %%@
different flavors in transversal magnetic field. In Sec.~\ref{CONCL} we summarize %%@
our results.

\section{Electrodynamics of mixed particle states possessing magnetic %%@
moments}\label{EDnu}

The magnetic moments of Dirac neutrinos are usually non-vanishing in both flavor %%@
and mass eigenstates bases~\cite{BabMat87}. This is why the flavor oscillations of %%@
neutrinos under the influence of an external magnetic field are associated %%@
generally with  spin-flavor conversions. One has to analyze the behaviour of a two %%@
Dirac neutrino system in a four dimensional basis %%@
$\Psi^\mathrm{T}=(\nu_\beta^\mathrm{L}, \nu_\alpha^\mathrm{L}, %%@
\nu_\beta^\mathrm{R}, \nu_\alpha^\mathrm{R})$,  which includes both  chiral %%@
components of neutrinos~\cite{GriSch93}. The analytic solution of the Schr\"odinger %%@
evolution equation for this system appears to be quite complicated generally.

Let us denote the magnetic moments of the two  Dirac neutrinos %%@
$\nu_{\alpha},\nu_{\beta}$ as ${M}_{\alpha\alpha}$, ${M}_{\beta\beta}$ and %%@
${M}_{\alpha\beta}$, where the last one is called as a transition magnetic moment. %%@
The Lagrangian of the neutrinos in the presence of an external electromagnetic %%@
field $F_{\mu\nu}=(\mathbf{E},\mathbf{B})$ is then given by
\begin{align}\label{Lagrnu}
  \mathcal{L}(\nu_{\alpha},\nu_{\beta})= &
  \sum_{\lambda=\alpha,\beta}\mathcal{L}_0(\nu_{\lambda})-
  (m_{{\beta}{\alpha}}\bar{\nu}_{\beta}\nu_{\alpha}+\mathrm{h.c.})
  \notag
  \\
  & -
  \frac{1}{2}
  \sum_{\lambda\lambda'=\alpha,\beta}
  M_{\lambda\lambda'}\bar{\nu}_{\lambda}\sigma_{\mu\nu}
  \nu_{\lambda'} F^{\mu\nu}.
\end{align}
Here, $\mathcal{L}_0(\nu_{\lambda})=\bar{\nu}_{\lambda}(\mathrm{i}\gamma^\mu %%@
\partial_\mu-m_{\lambda\lambda})\nu_{\lambda}$ and %%@
$\sigma_{\mu\nu}=(\mathrm{i}/2)(\gamma_\mu\gamma_\nu-\gamma_\nu\gamma_\mu)$, and %%@
the parameters $m_{\lambda\lambda} $ and $m_{\beta\alpha}$ (a vacuum mixing %%@
parameter) have dimension of mass.

To analyze the dynamics of the system one has to set the initial condition by %%@
giving the initial wave functions of neutrinos $\nu_{\lambda}$ %%@
($\lambda=\alpha,\beta$) and then analytically determine the field distributions at %%@
later moments of time. The analogous problem was studied in Refs.~\cite{FOvac} for %%@
mixed neutrino states  in vacuum and in Ref.~\cite{Dvo06EPJC} for mixed neutrino %%@
states in an external axial-vector field.

We assume an initial condition of the form
\begin{equation}\label{inicondnu}
  \nu_{\alpha}(\mathbf{r},0)=0,
  \quad
  \nu_{\beta}(\mathbf{r},0)=\xi(\mathbf{r}),
\end{equation}
where $\xi(\mathbf{r})$ is a function to be specified. If we identify %%@
$\nu_{\alpha}$ as $\nu_e$ or $\nu_\tau$ and $\nu_{\beta}$ as $\nu_{\mu}$, for %%@
example, this initial condition might correspond a situation where the source of %%@
neutrinos consists of pions and kaons which decay into muon neutrinos.

In order to eliminate the vacuum mixing term in Eq.~\eqref{Lagrnu} we introduce a %%@
new basis of the wave functions, the mass eigenstate basis $\psi_a$, $a=1,2$, %%@
related to the original flavor basis $\nu_{\lambda}$ through a unitary %%@
transformation
\begin{equation}\label{matrtrans}
  \nu_{\lambda}=\sum_{a=1,2}U_{\lambda a}\psi_a,
\end{equation}
where the matrix $U=({U}_{\lambda a})$ is parameterized in terms of a mixing angle %%@
$\theta$ in the usual manner:
\begin{equation}\label{matrU}
 U=
  \begin{pmatrix}
    \cos \theta & -\sin \theta \\
    \sin \theta & \cos \theta \
  \end{pmatrix}.
\end{equation}
The Lagrangian~\eqref{Lagrnu} rewritten in terms of the fields $\psi_a$ takes the %%@
form
\begin{align}\label{Lagrpsi}
  \mathcal{L}(\psi_1,\psi_2)= & \sum_{a=1,2}\mathcal{L}_0(\psi_a)
  \notag
  \\
  & -
  \frac{1}{2}
  \sum_{ab=1,2}\mu_{ab}\bar{\psi}_a\sigma_{\mu\nu}\psi_b F^{\mu\nu},
\end{align}
where $\mathcal{L}_0(\psi_a)=\bar{\psi}_a(\mathrm{i}\gamma^\mu %%@
\partial_\mu-m_a)\psi_a$ is the Lagrangian for the free neutrino $\psi_a$ with the %%@
mass $m_a$ and
\begin{equation}\label{magmomme}
  \mu_{ab}=
  \sum_{\lambda\lambda'=\alpha,\beta}
  U^{-1}_{a\lambda}{M}_{\lambda\lambda'}U_{\lambda' b},
\end{equation}
is the magnetic moment matrix presented in the mass eigenstates basis. Using %%@
Eqs.~\eqref{inicondnu}-\eqref{matrU}  the initial conditions for the fermions %%@
$\psi_a$ become
\begin{equation}\label{inicondpsi}
  \psi_1(\mathbf{r},0)=\sin\theta\xi(\mathbf{r}),
  \quad
  \psi_2(\mathbf{r},0)=\cos\theta\xi(\mathbf{r}).
\end{equation}

Let us assume that the magnetic field is constant, uniform and directed along the %%@
$z$-axis, $\mathbf{B}=(0, 0, B)$, and that the electric field vanishes, %%@
$\mathbf{E}=0$. In this case we write down the Pauli-Dirac equation for $\psi_a$, %%@
resulting from Eq.~\eqref{Lagrpsi}, as follows:
\begin{equation}\label{Direqpsi}
  \mathrm{i}\dot{\psi}_a=\mathcal{H}_a\psi_a+V\psi_b,
  \quad
  a,b=1,2,
  \quad
  a \neq b,
\end{equation}
where $\mathcal{H}_a=(\bm{\alpha}\mathbf{p})+\rho_3 m_a-\mu_a \rho_3 \Sigma_3 B$ is %%@
the Hamiltonian for the neutrino $\psi_a$ accounting for the magnetic field, %%@
$V=-\mu \rho_3 \Sigma_3 B$ describes the interaction of the transition magnetic %%@
moment with the external magnetic field, $\mu_a=\mu_{aa}$, and %%@
$\mu=\mu_{ab}=\mu_{ba}$ are elements of the matrix $({\mu}_{ab})$. Here we use the %%@
usual definitions for the Dirac matrixes $\bm{\alpha}=\gamma^0\bm{\gamma}$, %%@
$\rho_3=\gamma^0$ and $\bm{\Sigma}=\gamma^0\gamma^5\bm{\gamma}$.

It is in order to remark that in Ref.~\cite{BorZhuTer88} the problem of neutrino %%@
with anomalous magnetic moment was considered in the framework of Dirac-Schwinger %%@
equation in an external magnetic field (Furry representation), which leads to a %%@
wave equation  that differs from our equation~\eqref{Direqpsi}. The two approaches
give, however, the same energy spectrum [see Eq.~\eqref{energy}].

The general solution to Eq.~\eqref{Direqpsi} can be presented as follows:
\begin{align}\label{GsolDPeq}
  \psi_{a}(\mathbf{r},t)= &
  \int \frac{\mathrm{d}^3\mathbf{p}}{(2\pi)^{3/2}}
  e^{\mathrm{i}\mathbf{p}\mathbf{r}}
  %\notag
  %\\
  %& \times
  \sum_{\zeta=\pm 1}
  \big[
    a_a^{(\zeta)}(t)u_a^{(\zeta)}\exp{(-\mathrm{i}E_a^{(\zeta)} t)}
	\notag
	\\
	& +
    b_a^{(\zeta)}(t)v_a^{(\zeta)}\exp{(+\mathrm{i}E_a^{(\zeta)} t)}
  \big],
\end{align}
The basis spinors $u_a^{(\zeta)}$ and $v_a^{(\zeta)}$, as well as the energy %%@
$E_a^{(\zeta)}$, as a function of the particle momentum are given in the %%@
Appendix~\ref{AppDPE}. Our main goal is to determine the  coefficients %%@
$a_a^{(\zeta)}$ and $b_a^{(\zeta)}$ consistent with both the initial %%@
conditions~\eqref{inicondpsi} and the evolution equation~\eqref{Direqpsi}. They are %%@
in general functions of time.

\section{Perturbative solution}\label{PT}

In this section we investigate the dynamics of the mixed neutrinos system in the %%@
case where the transition magnetic moment can be considered as a small %%@
perturbation.  We assume that the characteristic energy of the particle $\psi_a$ is %%@
large compared with the energy $|\mu B|$ associated with the interaction induced by %%@
the transition magnetic moment.  Imposing the initial condition~\eqref{inicondpsi}, %%@
we now solve Eq.~\eqref{Direqpsi} taking the term  $V\psi_b$ in the %%@
Hamiltonian~\eqref{Direqpsi} as a small correction. We express the solution of %%@
Eq.~\eqref{Direqpsi} as an expansion
\begin{equation}\label{expan}
  \psi_{a}(\mathbf{r},t)=
  \psi_{a}^{(0)}(\mathbf{r},t)+\psi_{a}^{(1)}(\mathbf{r},t)+\dots,
\end{equation}
where $\psi_{a}^{(0)}(\mathbf{r},t)$ is the solution of Eq.~\eqref{Direqpsi} when %%@
$V\psi_b=0$, that is, the eigenvalue of the Hamiltonian $\mathcal{H}_a$. The %%@
function $\psi_{a}^{(1)}(\mathbf{r},t)$ is  linear in $\mu B$, and the ellipses %%@
stands for terms of higher order in $\mu B$.

Let us first search for the zeroth order solution $\psi_{a}^{(0)}$. The %%@
coefficients $a_a^{(\zeta)}$ and $b_a^{(\zeta)}$ defined in  Eq.~\eqref{GsolDPeq} %%@
are for $\psi_{a}^{(0)}$ time independent. Using the orthonormality conditions for %%@
the spinors $u_a^{(\zeta)}$ and
$v_a^{(\zeta)}$ [see Eq.~\eqref{ortcond}] and following  the
method developed in Refs.~\cite{FOvac}, we find
\begin{equation}\label{solpsi0}
  \psi_{a}^{(0)}(\mathbf{r},t)=
  \int \frac{\mathrm{d}^3\mathbf{p}}{(2\pi)^3}
  e^{\mathrm{i}\mathbf{p}\mathbf{r}}
  S_a(\mathbf{p},t)\psi_{a}(\mathbf{p},0),
\end{equation}
where
\begin{align}\label{PJfB}
  S_a(\mathbf{p},t)= &
  \sum_{\zeta=\pm 1}
  \Big[
    \left(
      u_a^{(\zeta)}\otimes u_a^{(\zeta)\dag}
    \right)
    \exp{(-\mathrm{i}E_a^{(\zeta)} t)}
	\notag
	\\
	& +
    \left(
      v_a^{(\zeta)}\otimes v_a^{(\zeta)\dag}
    \right)
    \exp{(+\mathrm{i}E_a^{(\zeta)} t)}
  \Big],
\end{align}
is an analog of the Pauli-Jordan function for a spin-$1/2$ particle in an external %%@
magnetic field and
\begin{equation*}
  \psi_{a}(\mathbf{p},0)=
  \int \mathrm{d}^3\mathbf{r}
  e^{-\mathrm{i}\mathbf{p}\mathbf{r}}
  \psi_{a}(\mathbf{r},0)
\end{equation*}
is the Fourier transform of the initial wave function~\eqref{inicondpsi}.

Eqs.~\eqref{solpsi0} and~\eqref{PJfB} together with Eqs.~\eqref{matrtrans} %%@
and~\eqref{inicondpsi} [or Eq.~\eqref{inicondnu}] allow one to describe the %%@
evolution of any system of two Dirac fermions in an external magnetic field. These %%@
expressions directly follow  from the Lorentz invariant Pauli-Dirac %%@
equation~\eqref{Direqpsi}, and they are valid for an arbitrary initial condition. %%@
In particular, the initial momentum of the particle can be arbitrary, though %%@
neutrinos are of course generally ultrarelativistic.

To proceed we have to specify our  initial condition, that is, give the function %%@
$\xi(\mathbf{r})$. Following our previous studies~\cite{FOvac}, we choose the %%@
initial condition for $\nu_\beta$ as a plane wave %%@
$\xi(\mathbf{r})=e^{\mathrm{i}\mathbf{k}\mathbf{r}}\xi_0$, with the initial %%@
momentum being directed along the $x$-axis, i.e.  $\mathbf{k}=(k, 0, 0)$. Given %%@
that we have chosen the magnetic field as $\mathbf{B}=(0, 0, B)$, we are thus %%@
studying the evolution of neutrino wavefunction in a transverse magnetic field.  We %%@
also consider neutrinos as ultrarelativistic particles, i.e. $k\gg m_{1,2}$, %%@
implying $\xi_0^\mathrm{T}=(1/2)(1, -1, -1, 1)$.  It is easy to see that $\xi_0$ %%@
obeys the condition $(1/2)(1-\Sigma_1)\xi_0=\xi_0$ and  is normalized to one. %%@
Hence, the spinor $\xi(\mathbf{r})$ describes an ultrarelativistic particle %%@
propagating along the $x$-axis with its spin directed opposite to the $x$-axis, %%@
i.e. a left-handedly polarized neutrino.

In contrast to the mixing due to a mass matrix, which is always present, the  %%@
magnetic transition only occurs in the presence of a magnetic field. Let us assume %%@
that a particle passes through a region of length $r_0$ where there is a %%@
nonvanishing magnetic field, i.e. $B\neq 0$ if $0\leq t\leq t_0$ and $B=0$ if $t<0$ %%@
and  $t>t_0$ where $t_0\sim r_0$. The particle emission and detection are assumed %%@
to take place while the magnetic field is switched off, at $t_e=0-\delta t$ and %%@
$t_d=t_0+\delta t$, respectively, where $\delta t$ is some small period of time. %%@
Hence, the operator $\Sigma_1$ describes the polarization state of the initial and %%@
final particle, and one can assume that the system is initially prepared to a state %%@
described in terms of the spinor $\xi_0$ as we have done above. Of course, the wave %%@
function must be continuous in the borderlines of  the $B=0$ and $B\neq 0$ regions.

One can now determine the field distribution of the particle $\nu_\alpha$, which %%@
was  initially absent. With help of Eqs.~\eqref{matrtrans}, \eqref{matrU}, %%@
\eqref{solpsi0} and~\eqref{PJfB}, and by taking into account the initial condition %%@
we have chosen, we obtain the following expression for the wavefunction of the %%@
right-polarized state of $\nu_\alpha$ :
\begin{align}\label{nu10}
  \nu_{\alpha}^{(0)\mathrm{R}}(x,t)= &
  \frac{1}{2}(1+\Sigma_1)
  \left[
    \cos\theta\psi_{1}^{(0)}(x,t)-\sin\theta\psi_{2}^{(0)}(x,t)
  \right]
  \notag
  \\
  = &
  \mathrm{i}\sin\theta\cos\theta
  \big[
    e^{-\mathrm{i}\mathcal{E}_1 t}\sin(\mu_1 Bt)
	\notag
	\\
	& -
    e^{-\mathrm{i}\mathcal{E}_2 t}\sin(\mu_2 Bt)
  \big]
  e^{\mathrm{i}kx}\kappa_0,
\end{align}%\nonumber
where
\begin{equation}\label{Ekappa}
  \mathcal{E}_{1,2}=\sqrt{m_{1,2}^2+k^2},
  \quad
  \kappa_0^\mathrm{T}=(1/2)(1, 1, 1, 1).
\end{equation}
We have used here the identities
\begin{gather}
  \label{noantpart}
  \left(
    v^{\pm{}} \otimes v^{\pm{}\dag}
  \right)\xi_0=0,
  \\
  \label{xi0k0}
  \frac{1}{2}
  (1+\Sigma_1)
  \left(
    u^{\pm{}} \otimes u^{\pm{}\dag}
  \right)\xi_0=
  \pm\frac{1}{2}\kappa_0,
\end{gather}
which result from Eq.~\eqref{urspinors}. Note that in the ultrarelativistic limit %%@
$m_a/k\ll 1$ one can neglect the neutrino mass dependence of the basis spinors %%@
$u^{(\zeta)}$ and $v^{(\zeta)}$ and this is why the subscripts $a$ and $b$ are %%@
omitted in Eq.~\eqref{xi0k0} (see also Sec.~\ref{BPT} and Eq.~\eqref{urspinors} for %%@
the explicit forms of $u^{(\zeta)}$ and $v^{(\zeta)}$). It should be mentioned that %%@
Eq.~\eqref{noantpart} implies that no particles with negative energies %%@
(antiparticles) are produced in neutrino interacting with an external magnetic %%@
field.

The measurable quantity is the square of the wave function %%@
$\nu_{\alpha}^{(0)\mathrm{R}}(x,t)$, which can be interpreted as the transition %%@
probability of the process $\nu_\beta^\mathrm{L}\to\nu_\alpha^\mathrm{R}$.  From %%@
Eq.~\eqref{nu10} we obtain
\begin{align}\label{Ptr0}
  P_{\nu_\beta^\mathrm{L}\to\nu_\alpha^\mathrm{R}}^{(0)}(t)= &
  \left|
    \nu_{\alpha}^{(0)\mathrm{R}}
  \right|^2
  =
  \sin^2(2\theta)
  \notag
  \big\{
    \sin^2(\delta\mu B t)\cos^2(\bar\mu B t)
	\\
	& +
    \sin(\mu_1 B t)\sin(\mu_2 B t)
    \sin^2
    \left[
      \Phi(k)t
    \right]
  \big\},
\end{align}
where
\begin{equation}\label{vacuumPh}
  \Phi(k)=\frac{\delta m^2}{4k},
\end{equation}
is the phase of vacuum oscillations, $\delta m^2 = m_1^2-m_2^2$,
$\delta\mu=(\mu_1-\mu_2)/2$ and $\bar\mu=(\mu_1+\mu_2)/2$.

In the case of neutrinos having equal magnetic moments, $\mu_1=\mu_2=\mu_0$,  %%@
Eq.~\eqref{Ptr0} leads to the result one would expect. Namely,  Eq.~\eqref{Ptr0} %%@
can be rewritten as $P=P_F P_S$, where $P_F = \sin^2(2\theta)\sin^2[\Phi(k)t]$ is %%@
the usual transition probability of flavor oscillation and $P_S = \sin^2(\mu_0 B %%@
t)$ is the probability of the transition between different polarization states %%@
within each mass eigenstate. That is, as the magnetic moment interactions are in %%@
this case insensitive to flavor, the transitions between flavors are solely due to %%@
the mass mixing.

Let us now consider the first order correction  $\psi_{a}^{(1)}(\mathbf{r},t)$  in %%@
Eq.~\eqref{expan}. Using the same method as in Ref.~\cite{Dvo06EPJC} we obtain
%
%\begin{widetext}
%
\begin{align}\label{solpsi1}
  \psi_{a}^{(1)}(\mathbf{r},t)= &
  -\mathrm{i}\int \frac{\mathrm{d}^3\mathbf{p}}{(2\pi)^3}
  e^{\mathrm{i}\mathbf{p}\mathbf{r}}
  \sum_{\zeta=\pm 1}
  \Big[
    \left(
      u_a^{(\zeta)}\otimes u_a^{(\zeta)\dag}
    \right)
	\notag
	\\
	& \times
    \exp{(-\mathrm{i}E_a^{(\zeta)} t)}
    V\mathcal{G}_a^\mathrm{(\zeta)}
    +
    \left(
      v_a^{(\zeta)}\otimes v_a^{(\zeta)\dag}
    \right)
	\notag
	\\
	& \times
    \exp{(+\mathrm{i}E_a^{(\zeta)} t)}
    V\mathcal{R}_a^\mathrm{(\zeta)}
  \Big]
  \psi_{b}(\mathbf{p},0),
\end{align}
%
%\end{widetext}
%
where
\begin{align}\label{PJfint}
  \mathcal{G}_a^{(\zeta)}= &
  \int_0^t \mathrm{d}t' \exp{(+\mathrm{i}E_a^{(\zeta)}t')}S_b(\mathbf{p},t'),
  \notag
  \\
  \mathcal{R}_a^{(\zeta)}= &
  \int_0^t \mathrm{d}t' \exp{(-\mathrm{i}E_a^{(\zeta)}t')}S_b(\mathbf{p},t').
\end{align}
In Eqs.~\eqref{solpsi1} and~\eqref{PJfint}  $a\neq b$.

With help of Eqs.~\eqref{matrtrans}, \eqref{matrU}, \eqref{solpsi1} %%@
and~\eqref{PJfint}  the first-order correction to $\nu_{\alpha}^{(0)}$ gets the %%@
form
\begin{align}\label{nu11}
  \nu_{\alpha}^{(1)\mathrm{R}}(x,t)= &
  \mathrm{i}\cos 2\theta
  \frac{\mu B}{2}
  \\
  \notag
  & \times
  \left(
    e^{-\mathrm{i}\Sigma t}\frac{\sin\Delta t}{\Delta}+
    e^{-\mathrm{i}\sigma t}\frac{\sin\delta t}{\delta}
  \right)
  e^{\mathrm{i}kx}\kappa_0,
\end{align}
where
\begin{gather}
  \notag
  \Sigma=\bar{\mathcal{E}}+\bar{\mu}B,
  \quad
  \Delta=\delta\mathcal{E}+\delta \mu B,
  \\
  \label{dsfun}
  \sigma=\bar{\mathcal{E}}-\bar{\mu}B,
  \quad
  \delta=\delta\mathcal{E}-\delta \mu B,
\end{gather}
and
\begin{equation}
  \delta\mathcal{E}=\frac{\mathcal{E}_1-\mathcal{E}_2}{2},
  \quad
  \bar{\mathcal{E}}=\frac{\mathcal{E}_1+\mathcal{E}_2}{2}.
\end{equation}
It can be seen from Eqs.~\eqref{nu11} and~\eqref{dsfun} that the additional %%@
constraint should be fulfilled for the perturbative approach to be valid, namely %%@
$\delta\mathcal{E} \neq \pm \delta \mu B$. It should be noticed that %%@
Eq.~\eqref{nu11} is obtained in the ultrarelativistic limit, where small terms %%@
$m_a/k\ll 1$ are neglected.

Let us now find out the effect of the first-order correction to the transition %%@
probability. Using Eqs.~\eqref{nu10} and~\eqref{nu11} we obtain
%
%\begin{widetext}
%
\begin{align}\label{Ptr1}
  P_{\nu_\beta^\mathrm{L}\to\nu_\alpha^\mathrm{R}}^{(1)}(t)= &
  \nu_{\alpha}^{(0)\mathrm{R}\dag}\nu_{\alpha}^{(1)\mathrm{R}}+\mathrm{h.c.}
  \\
  = &
  \frac{\mu B}{4[\Phi^2(k)-(\delta\mu B)^2]}
  \sin 4\theta
  \notag
  \\
  & \times
  \Big\{
    \Phi(k)
    \sin
    \left[
      2\Phi(k)t
    \right]
    \sin(2\delta\mu B t)
    \notag
    \\
    & -
    4\delta\mu B
    \big(
      \sin^2(\delta\mu B t)\cos^2(\bar\mu B t)
	  \notag
	  \\
	  \notag
	  & +
      \sin(\mu_1 B t)\sin(\mu_2 B t)
      \sin^2
      \left[
        \Phi(k)t
      \right]
    \big)
  \Big\}.
\end{align}
%
%\end{widetext}

The transition probability of spin-flavor oscillations between mass eigenstate %%@
neutrinos up to effects linear in the transition magnetic moment is given as a sum %%@
of the probabilities given in Eqs.~\eqref{Ptr0} and~\eqref{Ptr1}.

It should be noticed that the method used above allows one to examine particles %%@
with arbitrary initial distributions $\xi(\mathbf{r})$  since  Eqs.~\eqref{solpsi0} %%@
and~\eqref{PJfB}, as well as Eqs.~\eqref{solpsi1} and~\eqref{PJfint}, are not %%@
restricted to any specific initial condition. Also, as we mentioned before, the %%@
initial momenta of particles is not restricted. This has relevance, of course,  %%@
only in the case the results are applied to other particles than neutrinos.  %%@
Nevertheless, our analysis is not totally general as we have assumed the transition %%@
magnetic moment small. It turns out, actually, that one can circumvent this %%@
restriction in the case of ultrarelativistic particles. That is, in the case of %%@
neutrinos one can solve, as we will do in the next Section, the Pauli-Dirac %%@
equation analytically for an arbitrary magnetic moment matrix.

\section{Evolution of ultrarelativistic particles with arbitrary magnetic moments %%@
matrix}\label{BPT}

In this section we study the influence of the transition magnetic $\mu$ on the %%@
evolution of the mixed neutrino system without assuming   $\mu$ to be small.

Using the orthonormality conditions of the basis spinors [see Eq.~\eqref{ortcond}] %%@
and the fact that $\pm E_a^{(\zeta)}$ are the eigenvalues of the Hamiltonian %%@
$\mathcal{H}_a$, $a=1,2$,
\begin{equation}
  \mathcal{H}_a u_a^{(\zeta)}= + E_a^{(\zeta)} u_a^{(\zeta)},
  \quad
  \mathcal{H}_a v_a^{(\zeta)}= - E_a^{(\zeta)} v_a^{(\zeta)},
\end{equation}
one obtains from Eq.~\eqref{GsolDPeq} the following ordinary differential equations %%@
for $a_a^{(\zeta)}(t)$ and $b_a^{(\zeta)}(t)$:
\begin{align}\label{abEqsGc}
  \mathrm{i}\dot{a}_a^{(\zeta)}= &
  \exp{(+\mathrm{i}E_a^{(\zeta)}t)}u_a^{(\zeta)\dag}
  V
  \notag
  \\
  & \times
  {\sum_{\zeta'=\pm 1}}
  \big[
    a_b^{(\zeta')}u_b^{(\zeta')}\exp{(-\mathrm{i}E_b^{(\zeta')} t)}
	\notag
	\\
	& +
    b_b^{(\zeta')}v_b^{(\zeta')}\exp{(+\mathrm{i}E_b^{(\zeta')} t)}
  \big],
  \notag
  \\
  \mathrm{i}\dot{b}_a^{(\zeta)}= &
  \exp{(-\mathrm{i}E_a^{(\zeta)} t)}v_a^{(\zeta)\dag}
  V
  \notag
  \\
  & \times
  {\sum_{\zeta'=\pm 1}}
  \big[
    a_b^{(\zeta')}u_b^{(\zeta')}\exp{(-\mathrm{i}E_b^{(\zeta')} t)}
	\notag
	\\
	& +
    b_b^{(\zeta')}v_b^{(\zeta')}\exp{(+\mathrm{i}E_b^{(\zeta')} t)}
  \big].
\end{align}
Eqs.~\eqref{abEqsGc} are  subject to the initial conditions
\begin{align}
  a_a^{(\zeta)}(0)= &
  \frac{1}{(2\pi)^{3/2}}u_a^{(\zeta)\dag}\psi_a(\mathbf{p},0),
  \notag
  \\
  b_a^{(\zeta)}(0)= &
  \frac{1}{(2\pi)^{3/2}}v_a^{(\zeta)\dag}\psi_a(\mathbf{p},0),
\end{align}
which result from Eq.~\eqref{GsolDPeq}. Note that the functions $a_a^{(\zeta)}$ and %%@
$b_a^{(\zeta)}$ are here  time dependent in general, in contrast to Sec.~\ref{PT}.

Let us choose the initial wave function of $\nu_\beta$ [see Eq.~\eqref{inicondnu}] %%@
as
$\xi(\mathbf{r})=e^{\mathrm{i}\mathbf{k}\mathbf{r}}\xi_0$, with the initial %%@
momentum being aligned along the $x$-axis, $\mathbf{k}=(k, 0, 0)$. The magnetic %%@
field is assumed to be $\mathbf{B}=(0, 0, B)$, implying that  we study the %%@
propagation of neutrinos in the transversal magnetic field. Given that %%@
$\mathbf{k}\perp\mathbf{B}$, we can rewrite the Eq.~\eqref{abEqsGc}  for the %%@
function  $a^{(\zeta)}_a$  in the form  (the corresponding equation  for the %%@
function $b^{(\zeta)}_a$ can be obtained analogously)
\begin{align}\label{aEqskpB}
  \mathrm{i}\dot{a}_a^{\pm{}}= &
  a_b^{\pm{}}
  \langle u_a^{\pm{}} | V | u_b^{\pm{}} \rangle
  \exp{[\mathrm{i}(E_a^{\pm{}}-E_b^{\pm{}}) t]}
  \notag
  \\
  & +
  b_b^{\mp{}}
  \langle u_a^{\pm{}} | V | v_b^{\mp{}} \rangle
  \exp{[\mathrm{i}(E_a^{\pm{}}+E_b^{\mp{}}) t]}.
\end{align}
Here we use the explicit form of the basis spinors in the transversal magnetic %%@
field given in Eq.~\eqref{spinors}. Note that  $\langle u_a^{\pm{}} | V | %%@
u_b^{\mp{}} \rangle = 0$ and $\langle u_a^{\pm{}} | V | v_b^{\pm{}} \rangle = 0$.

A further simplification of  Eqs.~\eqref{abEqsGc} is obtained when we study the %%@
ultrarelativistic initial wave function, $k\gg m_{1,2}$, and %%@
$\xi_0^\mathrm{T}=(1/2)(1, -1, -1, 1)$ (see Sec.~\ref{PT}), in other words the %%@
system is, like in the case we studied in the previous Section, in the state %%@
$\nu_\beta^\mathrm{L}$ initially. With help of the obvious identities $\langle %%@
u_a^{\pm{}} | V | u_b^{\pm{}} \rangle=\mp \mu B$ and $\langle u_a^{\pm{}} | V | %%@
v_b^{\mp{}} \rangle=0$, which result from Eq.~\eqref{urspinors}, one can cast %%@
Eq.~\eqref{aEqskpB} into the form
\begin{equation}\label{aEqsff}
  \mathrm{i}\dot{a}_a^{\pm{}}=
  \mp a_b^{\pm{}}\mu B
  \exp{[\mathrm{i}(E_a^{\pm{}}-E_b^{\pm{}}) t]}.
\end{equation}
Let us note that the Eq.~\eqref{aEqsff} is similar to the evolution equations for a %%@
neutrino interacting with a twisting magnetic field, which were examined in %%@
Refs.~\cite{twisting}. On the basis of this previous study we are able to write %%@
down the solutions as
\begin{align}\label{asol}
  a_1^{\pm{}}(t)= &
  F^{\pm{}}a_1^{\pm{}}(0)+G^{\pm{}}a_2^{\pm{}}(0),
  \notag
  \\
  a_2^{\pm{}}(t)= &
  F^{\pm{}*{}}a_2^{\pm{}}(0)-G^{\pm{}*{}}a_1^{\pm{}}(0),
\end{align}
where
\begin{align}\label{fg}
  F^{\pm{}}= &
  \left[
    \cos\Omega_{\pm{}}t-
    \mathrm{i}\frac{\omega_{\pm{}}}{2\Omega_{\pm{}}}\sin\Omega_{\pm{}}t
  \right]\exp{(\mathrm{i}\omega_{\pm{}}t/2)},
  \notag
  \\
  G^{\pm{}}= &
  \pm\mathrm{i}\frac{\mu B}{\Omega_{\pm{}}}\sin\Omega_{\pm{}}t
  \exp{(\mathrm{i}\omega_{\pm{}}t/2)},
\end{align}
and
\begin{equation}\label{Omegaomega}
  \Omega_{\pm{}}=\sqrt{(\mu B)^2+(\omega_{\pm{}}/2)^2},
  \quad
  \omega_{\pm{}}=E_1^{\pm{}}-E_2^{\pm{}}.
\end{equation}
The  derivation of Eqs.~\eqref{asol}-\eqref{Omegaomega} from Eqs.~\eqref{aEqsff} is %%@
presented in Appendix~\ref{detailsabsol}.

Using Eq.~\eqref{GsolDPeq} and Eqs.~\eqref{abEqsGc}-\eqref{Omegaomega} and the %%@
identity $\left( v^{(\zeta)} \otimes v^{(\zeta)\dag} \right)\xi_0=0$ [see %%@
Eq.~\eqref{xi0k0}] we
obtain the wave functions  $\psi_a$, $a=1,2$, as,
\begin{align}\label{psisolff}
  \psi_1(x,t)= &
  \exp{(-\mathrm{i}E_1^{+{}}t)}
  \left(
  u^{+{}}\otimes u^{+{}\dag}
  \right)
  \notag
  \\
  & \times
  [F^{+{}}\psi_1(x,0)+G^{+{}}\psi_2(x,0)]
  \notag
  \\
  & +
  \exp{(-\mathrm{i}E_1^{-{}}t)}
  \left(
    u^{-{}}\otimes u^{-{}\dag}
  \right)
  \notag
  \\
  & \times
  [F^{-{}}\psi_1(x,0)+G^{-{}}\psi_2(x,0)],
  \notag
  \\
  \psi_2(x,t)= &
  \exp{(-\mathrm{i}E_2^{+{}}t)}
  \left(
    u^{+{}}\otimes u^{+{}\dag}
  \right)
  \notag
  \\
  & \times
  [F^{+{}*{}}\psi_2(x,0)-G^{+{}*{}}\psi_1(x,0)]
  \notag
  \\
  & +
  \exp{(-\mathrm{i}E_2^{-{}}t)}
  \left(
    u^{-{}}\otimes u^{-{}\dag}
  \right)
  \notag
  \\
  & \times
  [F^{-{}*{}}\psi_2(x,0)-G^{-{}*{}}\psi_1(x,0)],
\end{align}
which satisfy the chosen initial condition since $G^{\pm{}}(0)=0$, $F^{\pm{}}(0)=1$ %%@
[see Eqs.~\eqref{fg}] and $[(u^{+{}}\otimes u^{+{}\dag})+(u^{-{}}\otimes %%@
u^{-{}\dag})]\psi_a(x,0) = \psi_a(x,0)$ [see Eq.~\eqref{urspinors}]. Note that the %%@
subscripts $a$ and $b$ are again omitted in the basis spinors $u^{(\zeta)}$ and %%@
$v^{(\zeta)}$ as we assume ultrarelativistic particles.

With help of Eqs.~\eqref{matrtrans}, \eqref{matrU} and~\eqref{psisolff} [see also %%@
Eq.~\eqref{nu10}] we receive for the right-handedly polarized component of %%@
$\nu_\alpha$ the expression
\begin{align}\label{nu1}
  \nu_\alpha^{\mathrm{R}}(x,t)= &
  \frac{1}{2}
  \big\{
    \sin\theta\cos\theta
    \big[
      e^{-\mathrm{i}\mathcal{E}_1 t}
      (e^{\mathrm{i}\mu_1 B t}F^{+{}}-e^{-\mathrm{i}\mu_1 B t}F^{-{}})
      \notag
      \\
      & -
      e^{-\mathrm{i}\mathcal{E}_2 t}
      (e^{\mathrm{i}\mu_2 B t}F^{+{}*{}}-e^{-\mathrm{i}\mu_2 B t}F^{-{}*{}})
    \big]
    \notag
    \\
    & +
    \cos^2\theta
    e^{-\mathrm{i}\mathcal{E}_1 t}
    (e^{\mathrm{i}\mu_1 B t}G^{+{}}-e^{-\mathrm{i}\mu_1 B t}G^{-{}})
    \notag
    \\
    & +
    \sin^2\theta
    e^{-\mathrm{i}\mathcal{E}_2 t}
    (e^{\mathrm{i}\mu_2 B t}G^{+{}*{}}-e^{-\mathrm{i}\mu_2 B t}G^{-{}*{}})
  \big\}
  \notag
  \\
  & \times
  e^{\mathrm{i}kx}\kappa_0.
\end{align}
The quantities $\mathcal{E}_a$ and $\kappa_0$ are defined in Eq.~\eqref{Ekappa}. We %%@
have used here the identity given in Eq.~\eqref{xi0k0}.

Finally, taking into account Eqs.~\eqref{fg} and~\eqref{Omegaomega} it is possible %%@
to express the wave function in Eq.~\eqref{nu1} in the  form
\begin{align}\label{nu1exp}
  \nu_\alpha^{\mathrm{R}}(x,t)= &
  \bigg\{
    \sin\theta\cos\theta
    \frac{1}{2\mathrm{i}}
    \bigg[
      \frac{\omega_{+{}}}{\Omega_{+{}}}\sin(\Omega_{+{}}t)
      \exp{(\mathrm{i}\bar{\mu}Bt)}
	  \notag
	  \\
	  & -
      \frac{\omega_{-{}}}{\Omega_{-{}}}\sin(\Omega_{-{}}t)
      \exp{(-\mathrm{i}\bar{\mu}Bt)}
    \bigg]
    \notag
    \\
    & +
    \mathrm{i}\mu B
    \left[
      \frac{\sin(\Omega_{+{}}t)}{\Omega_{+{}}}\cos^2\theta-
      \frac{\sin(\Omega_{-{}}t)}{\Omega_{-{}}}\sin^2\theta
    \right]
	\notag
	\\
	& \times
    \cos(\bar{\mu}Bt)
  \bigg\}
  \exp{(-\mathrm{i}\bar{\mathcal{E}}t+\mathrm{i}kx)}\kappa_0.
\end{align}
The transition probability for the process %%@
$\nu_\beta^\mathrm{L}\to\nu_\alpha^\mathrm{R}$ can be directly obtained as the  %%@
squared modulus of $\nu_\alpha^{\mathrm{R}}(x,t)$ from Eq.~\eqref{nu1} or %%@
Eq.~\eqref{nu1exp}, that is  $P_{\nu_\beta^\mathrm{L}\to\nu_\alpha^\mathrm{R}}(t)= %%@
|\nu_\alpha^{\mathrm{R}}(x,t)|^2$. Notice that the probability is a function of %%@
time alone with no dependence  on spatial coordinates. This is of course obvious as %%@
we have taken the initial wave function as a plane wave and the the magnetic field %%@
spatially constant.

Let us now apply the general results Eq.~\eqref{nu1} or Eq.~\eqref{nu1exp} to two %%@
special cases. We first consider the situation where $\mu_{1,2}\gg\mu$, i.e. the %%@
case when the transition magnetic moment is small compared with the diagonal ones. %%@
Using Eqs.~\eqref{fg} and~\eqref{Omegaomega} we find that in this case %%@
$F^{\pm{}}\approx 1$ and $\Omega_{\pm{}}\approx\omega_{\pm{}}/2$, and  %%@
Eq.~\eqref{nu1} takes the form
\begin{align}\label{nu1pt}
  \nu_\alpha^{\mathrm{R}}(x,t)\approx &
  \mathrm{i}
  \bigg\{
    \sin\theta\cos\theta
    \left[
      e^{-\mathrm{i}\mathcal{E}_1 t}\sin\mu_1 B t-
      e^{-\mathrm{i}\mathcal{E}_2 t}\sin\mu_2 B t
    \right]
    \notag
    \\
    & +
    \cos 2\theta\frac{\mu B}{2}
    \left(
      e^{-\mathrm{i}\Sigma t}\frac{\sin\Delta t}{\Delta}+
      e^{-\mathrm{i}\sigma t}\frac{\sin\delta t}{\delta}
    \right)
  \bigg\}
  \notag
  \\
  & \times
  e^{\mathrm{i}kx}\kappa_0.
\end{align}
In the previous Section we studied this case perturbatively, and one can easily %%@
check that results obtained there [see Eqs.~\eqref{nu10} and~\eqref{nu11}] coincide %%@
with (\ref{nu1pt}).

\begin{figure}
  \centering
  \includegraphics[scale=.41]{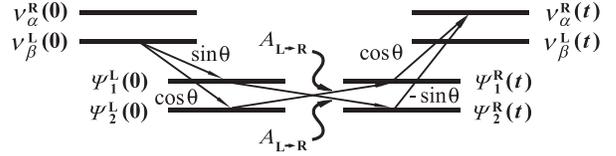}
  \caption{\label{diagram}
  The schematic illustration of the system evolution in the case
  $\mu\gg\mu_{1,2}$. The horizontal lines of the figure correspond to
  various neutrino eigenstates at different
  moments of time ($t=0$ and $t$). 
  The expressions next to arrows correspond to the
  appropriate factors in the formula~\eqref{nu1bpt} of the
  wave function. The arrows from $\nu_\beta^\mathrm{L}(0)$ to
  $\psi_1^\mathrm{L}(0)$ and
  $\psi_2^\mathrm{L}(0)$, for example, indicate the vacuum mixing matrix
  transformation at $t=0$ and the
  arrow from $\psi_1^\mathrm{L}(0)$ to $\psi_2^\mathrm{R}(t)$
  the evolution of the mass eigenstates with the
  helicity change.
  The transitions $\psi_a^\mathrm{L}(0)\to\psi_b^\mathrm{R}(t)$ 
  can be described by the formula,
  $\psi_{1,2}^{\mathrm{R}}(t) =
  A_{\mathrm{L}\to\mathrm{R}}\psi_{2,1}^{\mathrm{L}}(0)$, where
  $A_{\mathrm{L}\to\mathrm{R}}=\mathrm{i}(\mu B/\Omega)\sin \Omega t$.}
\end{figure}

As another application of our general result we will study the situation, where the %%@
transition magnetic moments are much larger than the diagonal ones, that is  %%@
$\mu\gg\mu_{1,2}$.  In this case Eqs.~\eqref{fg} gives $F^{+{}}\approx F^{-{}}$ and %%@
$G^{+{}}\approx - G^{-{}}$, and we receive from Eq.~\eqref{nu1} or  %%@
Eq.~\eqref{nu1exp} for the wave function $\nu_\alpha^{\mathrm{R}}$ the expression
\begin{align}\label{nu1bpt}
  \nu_\alpha^{\mathrm{R}}(x,t)\approx &
  \mathrm{i}\exp{(-\mathrm{i}\bar{\mathcal{E}}t+\mathrm{i}kx)}
  \notag
  \\
  & \times
  \cos (2\theta)
  \frac{\mu B}{\Omega}\sin(\Omega t) \kappa_0,
\end{align}
where
\begin{equation}\label{OmegaMaj}
  \Omega=\sqrt{(\mu B)^2+\Phi^2(k)}.
\end{equation}
The transition probability for the process %%@
$\nu_\beta^\mathrm{L}\to\nu_\alpha^\mathrm{R}$ is then given by
\begin{equation}\label{Ptrbpt}
  P_{\nu_\beta^\mathrm{L}\to\nu_\alpha^\mathrm{R}}(t)=
  \cos^2(2\theta)
  \left(
    \frac{\mu B}{\Omega}
  \right)^2
  \sin^2(\Omega t).
\end{equation}
The behavior of the system in this case is schematically
illustrated in Fig.~\ref{diagram}. It should be noticed that the analog of %%@
Eq.~\eqref{Ptrbpt} was obtained in Ref.~\cite{LimMar88} where the authors studied %%@
the resonant spin-flavor precession of Dirac and Majorana neutrinos in matter under %%@
the influence of an external magnetic field in frames of the quantum mechanical %%@
approach.

Let us discuss the applicability of our results to one specific oscillation  %%@
channel, $\nu_\mu^\mathrm{L}\xrightarrow{B}\nu_\tau^\mathrm{R}$.
According to the recent experimental data (see, e.g., Ref.~\cite{MalSchTorVal04}) %%@
the mixing angle between $\nu_\mu$ and $\nu_\tau$ is close to its maximal value of  %%@
$\pi/4$. In the limit of maximal mixing the magnetic moment matrix given in %%@
Eq.~\eqref{magmomme} takes the form
%
%\begin{widetext}
%
\begin{multline}\label{magmommepi4}
  ({\mu}_{ab})
  \approx
  \\
  \begin{pmatrix}
    ({M}_{\tau\tau}+{M}_{\mu\mu})/2+{M}_{\tau\mu} &
    -({M}_{\tau\tau}-{M}_{\mu\mu})/2 \\
    -({M}_{\tau\tau}-{M}_{\mu\mu})/2 &
    ({M}_{\tau\tau}+{M}_{\mu\mu})/2-{M}_{\tau\mu} \
  \end{pmatrix}.
\end{multline}
%
%\end{widetext}

Eqs.~\eqref{Ptr0} and~\eqref{Ptr1} [or Eq.~\eqref{nu1pt}] are valid in the case %%@
this matrix is close to diagonal, i.e. when  %%@
$|({M}_{\tau\tau}-{M}_{\mu\mu})/2|\ll|({M}_{\tau\tau}+{M}_{\mu\mu})/2 %%@
\pm{M}_{\tau\mu}|$.  In contrast to the mixing angles, the experimental information %%@
about the neutrino magnetic moments is very limited. Also theoretically, very %%@
little is known about their magnitude.  What one knows is that the diagonal %%@
magnetic moments $M_{\lambda\lambda}$ could be very small in the extensions of the %%@
Standard Model ${M}_{\lambda\lambda}\sim %%@
10^{-19}({m}_{\lambda\lambda}/\text{eV})\mu_\mathrm{B}$ (see, e.g., %%@
Refs.~\cite{nuMM}). The contributions to the transition magnetic moments, %%@
${M}_{\tau\mu}$ in our case, can be much bigger close to the experimental upper %%@
limit of $10^{-10}\mu_\mathrm{B}$~\cite{Yao06}. One can see that for any %%@
conceivable values of the masses of the known neutrinos,
${M}_{\mu\mu}$ and ${M}_{\tau\tau}$ are orders of magnitude smaller than %%@
$10^{-10}\mu_\mathrm{B}$. Our results [Eq.~\eqref{Ptr0}] are  applicable in this %%@
kind of  situation.

We can also consider  the opposite situation of $M_{\lambda\lambda}\gg %%@
{M}_{\tau\mu}$. It is implemented, e.g., in the minimally extended standard model %%@
(see Refs.~\cite{nuMM}) when the magnetic moments matrix is diagonal in the flavor %%@
basis, i.e. ${M}_{\tau\mu}$ is negligible. If it is the case, one observes that %%@
$\mu_1=\mu_2$ in Eq.~\eqref{magmommepi4}. Thus $\omega_{+{}}=\omega_{-{}}=\omega$ %%@
and $\Omega_{+{}}=\Omega_{-{}}=\Omega$ [see Eqs.~\eqref{Omegaomega} %%@
and~\eqref{urenergy}]. Eq.~\eqref{nu1exp} takes the form,
\begin{align}\label{nu1pi4}
  \nu_\alpha^{\mathrm{R}}(x,t)= &
  \frac{\omega}{2\Omega}\sin(\Omega t)\sin(\bar{\mu}B t)
  \notag
  \\
  & \times
  \exp{(-\mathrm{i}\bar{\mathcal{E}}t+\mathrm{i}kx)}\kappa_0,
\end{align}
where we account for that $\cos\theta=\sin\theta=1/\sqrt{2}$. The transition %%@
probability can be calculated on the basis of Eq.~\eqref{nu1pi4},
\begin{equation}\label{Ptrpi4}
  P_{\nu_\beta^\mathrm{L}\to\nu_\alpha^\mathrm{R}}(t)=
  \left[
    \frac{\Phi(k)}{\Omega}
  \right]^2
  \sin^2(\Omega t)\sin^2(\bar{\mu}B t).
\end{equation}
Now the magnetic moments are
$\bar{\mu}=({M}_{\tau\tau}+{M}_{\mu\mu})/2$ and
$\mu=-({M}_{\tau\tau}-{M}_{\mu\mu})/2$. In Eq.~\eqref{Ptrpi4} the
phase of vacuum oscillations, $\Phi(k)$, is determined in
Eq.~\eqref{vacuumPh} and the parameter $\Omega$ is introduced in
Eq.~\eqref{OmegaMaj}.

At the end of this section it should be noticed that the our results for the %%@
description of spin-flavor oscillations of relativistic Dirac neutrinos are %%@
consistent with the standard quantum mechanical approach based on the Schr\"odinger %%@
equation of evolution. The consistency is briefly discussed in Appendix~\ref{qmd}.    

\section{Summary}\label{CONCL}

We have studied in the framework of relativistic quantum mechanics the evolution of %%@
a system of two Dirac neutrinos that mix with each other and have non-vanishing %%@
magnetic moments. By solving the Pauli-Dirac equation with a given initial %%@
condition we determined the time evolution  of this system in an external magnetic %%@
field. We applied the recently developed approach for the description of the first %%@
quantized mixed particles (see Refs.~\cite{FOvac,Dvo06EPJC}).

We first (Sec.~\ref{PT}) studied the special case of the magnetic moments matrix, %%@
which is close to diagonal in the mass eigenstates basis. We obtained the general %%@
expressions for the fermion fields distributions exactly accounting for an external %%@
magnetic field in the leading and the next-to-leading orders in the transition %%@
magnetic moment in the mass eigenstates basis. As we started with the Lorentz %%@
invariant neutrino wave equations, the derived neutrino wave functions are valid %%@
for arbitrary initial conditions. For instance, with help of Eqs.~\eqref{solpsi0} %%@
and~\eqref{solpsi1} one can describe the propagation of non-relativistic particles %%@
in an external magnetic field.

We considered spin-flavor oscillations of ultrarelativistic neutrinos in a constant %%@
transversal magnetic field and  derived an analytical expression for the transition %%@
probability for the spin-flavor conversion in the limit where the energy associated %%@
with the magnetic interaction arising from transition magnetic moment is small %%@
compared with the total neutrino energy. Then (Sec.~\ref{BPT}) we examined the %%@
evolution of ultrarelativistic Dirac neutrinos in the constant transversal magnetic %%@
field. In this situation it was possible to take into account the influence of all %%@
magnetic moments on the neutrinos evolution, in contrast to the analysis in %%@
Sec.~\ref{PT} where the effect of transition magnetic moments where considered as a %%@
small perturbation. 

Our results are valid for  arbitrary magnetic field strength. The general %%@
analytical formula~\eqref{nu1exp} for the neutrino wave function allows one to %%@
calculate the transition probability and consider various limiting cases. To our %%@
knowledge,  the general case of spin-flavor oscillations of Dirac neutrinos with an %%@
arbitrary magnetic moment matrix has not been studied analytically before. Both %%@
experimental and theoretical information about magnetic moments of Dirac neutrinos %%@
is still very limited (see, i.e. Ref.~\cite{Yao06}). In this light  our results may %%@
turn out useful as they allow one to describe phenomenologically spin-flavor %%@
oscillations of Dirac neutrinos in any model which predicts neutrino magnetic %%@
moments. Although we  consider neutrinos, our formalism is straightforwardly %%@
applicable to any spin-1/2 particles.

\begin{acknowledgments}
The work has been supported by the Academy of Finland under the
contracts No.~108875 and No.~104915.
MD is thankful to the Russian Science Support Foundation for a grant as well as to %%@
Anatoly Borisov, Andrey Lobanov (MSU) and Timur Rashba (MPI, Munich) for helpful %%@
discussions. A special thank is addressed to Alexey Nemkin for his help on the %%@
illustration preparation. The referees' comments are also appreciated.  
\end{acknowledgments}

\appendix

\section{Solution to the Pauli-Dirac equation\label{AppDPE}}

The general solution to the Pauli-Dirac equation of the form [see also %%@
Eq.~\eqref{Direqpsi}], $\mathrm{i}\dot{\psi}_a=\mathcal{H}_a\psi_a$, for a massive %%@
neutral fermion with the anomalous magnetic moment was found in %%@
Ref.~\cite{TerBagKha65}. The energy as a function of the particle momentum %%@
$\mathbf{p}=(p_1,p_2,p_3)$ and the spin quantum number $\zeta=\pm 1$, which %%@
characterizes the spin direction with respect to the magnetic field, has the form
\begin{align}\label{energy}
  E_a^{(\zeta)} = & \sqrt{p_3^2+\mathcal{E}_a^{(\zeta)2}},
  \notag
  \\
  \mathcal{E}_a^{(\zeta)} = & \mathcal{E}_a-\zeta \mu_a B,
  \quad
  \mathcal{E}_a = \sqrt{m_a^2+p_1^2+p_2^2}.
\end{align}
The basis spinors are expressed in following way:
\begin{align}\label{spinors}
  u_a^{(\zeta)}= &
  \frac{1}{2\sqrt{E_a^{(\zeta)}}}
  \begin{pmatrix}
     \phi^{+{}}_a \alpha^{+{}}_a \\
     -\zeta\phi^{-{}}_a \alpha^{-{}}_a e^{\mathrm{i}\varphi} \\
     \phi^{+{}}_a \alpha^{-{}}_a \\
     \zeta\phi^{-{}}_a \alpha^{+{}}_a e^{\mathrm{i}\varphi} \
  \end{pmatrix},
  \\
  v_a^{(\zeta)}= &
  \frac{1}{2\sqrt{E_a^{(\zeta)}}}
  \begin{pmatrix}
     \phi^{+{}}_a \alpha^{-{}}_a \\
     \zeta\phi^{-{}}_a \alpha^{+{}}_a e^{\mathrm{i}\varphi} \\
     -\phi^{+{}}_a \alpha^{+{}}_a \\
     \zeta\phi^{-{}}_a \alpha^{-{}}_a e^{\mathrm{i}\varphi} \
  \end{pmatrix},
\end{align}
where
\begin{equation*}
  \phi^{\pm{}}_a=\sqrt{1\pm\zeta m_a/\mathcal{E}_a},
  \quad
  \alpha^{\pm{}}_a=\sqrt{E_a^{(\zeta)}\pm\zeta\mathcal{E}_a^{(\zeta)}},
\end{equation*}
and $\tan \varphi = p_2/p_1$. The quantum number $\zeta=\pm 1$ characterizes the %%@
direction of the particle spin with respect to the magnetic field since the %%@
polarization operator
\begin{equation*}
  \Pi_a = m_a \Sigma_3+\rho_2[\bm{\Sigma}\times\mathbf{p}]_3-\mu_a B,
\end{equation*}
where $\rho_2=\mathrm{i}\gamma^0\gamma^5$, commutes with the Hamiltonian %%@
$\mathcal{H}_a$. We also present the orthonormality conditions for the spinors %%@
$u_a^{(\zeta)}$ and $v_a^{(\zeta)}$,
\begin{equation}\label{ortcond}
  u_a^{(\zeta)\dag}u_a^{(\zeta')}=v_a^{(\zeta)\dag}v_a^{(\zeta')}=
  \delta_{\zeta\zeta'},
  \quad
  u_a^{(\zeta)\dag}v_a^{(\zeta')}=0,
\end{equation}
which follow from Eq.~\eqref{spinors}. The details of the derivation of %%@
Eqs.~\eqref{energy} and~\eqref{spinors} can be found in Ref.~\cite{TerBagKha65}.

If we study the propagation of ultrarelativistic particles in a transversal %%@
magnetic field, Eqs.~\eqref{energy} and~\eqref{spinors} take the form
\begin{equation}\label{urenergy}
  E_a^{(\zeta)} = \mathcal{E}_a-\zeta \mu_a B
  \approx p + \frac{m_a^2}{2p} - \zeta \mu_a B,
\end{equation}
and
\begin{align}\label{urspinors}
  u^{+{}}= &
  \frac{1}{\sqrt{2}}
  \begin{pmatrix}
     1 \\
     0 \\
     0 \\
     1 \
  \end{pmatrix},
  \quad
  u^{-{}}=
  \frac{1}{\sqrt{2}}
  \begin{pmatrix}
     0 \\
     1 \\
     1 \\
     0 \
  \end{pmatrix},
  \notag
  \\
  v^{+{}}= &
  \frac{1}{\sqrt{2}}
  \begin{pmatrix}
     0 \\
     1 \\
     -1 \\
     0 \
  \end{pmatrix},
  \quad
  v^{-{}}= 
  \frac{1}{\sqrt{2}}
  \begin{pmatrix}
     1 \\
     0 \\
     0 \\
     -1 \
  \end{pmatrix}.
\end{align}
In Eqs.~\eqref{urenergy} and~\eqref{urspinors} we assume that the momentum is %%@
directed along the $x$-axis, $\mathbf{p}=(p,0,0)$. The subscript $a$ is omitted in %%@
Eq.~\eqref{urspinors} since we neglect small terms $(m_a/p)\ll 1$ there.

\section{Solution to the ordinary differential equations for the functions %%@
$a_a^{(\zeta)}$ and $b_a^{(\zeta)}$\label{detailsabsol}}

Let us study the time evolution of the two component spinor %%@
$\mathbf{Z}^\mathrm{T}=(Z_1,Z_2)$ which is governed by the Schr\"odinger equation %%@
of the form,
\begin{equation}\label{Zequ}
  \mathrm{i}\dot{\mathbf{Z}}=H\mathbf{Z},
\end{equation}
where the Hamiltonian has the following form:
\begin{equation}\label{ZHam}
  H=\mu\mathcal{B}
  \begin{pmatrix}
    0 & e^{\mathrm{i}\omega t} \\
    e^{-\mathrm{i}\omega t} & 0 \
  \end{pmatrix}.
\end{equation}
Here $\mu$, $\mathcal{B}$ and $\omega$ are the real parameters. Eq.~\eqref{Zequ} %%@
should be supplied with the initial condition $\mathbf{Z}(0)$. To find the solution %%@
to Eqs.~\eqref{Zequ} and~\eqref{ZHam} we introduce the new spinor $\mathbf{Z}'$ by %%@
the relation, $\mathbf{Z}=\mathcal{U}\mathbf{Z}'$, where the unitary matrix %%@
$\mathcal{U}$ reads
\begin{equation}\label{ZZ'matr}
  \mathcal{U}=
  \begin{pmatrix}
    e^{\mathrm{i}\omega t/2} & 0 \\
    0 & e^{-\mathrm{i}\omega t/2} \
  \end{pmatrix}.
\end{equation}
Now Eq.~\eqref{Zequ} is rewritten in the following way:
\begin{equation}\label{Z'equ}
  \mathrm{i}\dot{\mathbf{Z}}'=H'\mathbf{Z}',
\end{equation}
with the new Hamiltonian $H'$ which is obtained with help of %%@
Eqs.~\eqref{Zequ}-\eqref{ZZ'matr},
\begin{equation}\label{Z'Ham}
  H'=\mathcal{U}^\dag H \mathcal{U} -
  \mathrm{i}\mathcal{U}^\dag\dot{\mathcal{U}}=
  \begin{pmatrix}
    \omega/2 & \mu\mathcal{B} \\
    \mu\mathcal{B} & -\omega/2 \
  \end{pmatrix}
\end{equation}
Note that the initial condition for the spinor $\mathbf{Z}'(0)$ is the same as for %%@
$\mathbf{Z}(0)$, $\mathbf{Z}'(0)=\mathbf{Z}(0)$, due to the special form of the %%@
matrix $\mathcal{U}$ in Eq.~\eqref{ZZ'matr}. 

Supposing that the Hamiltonian $H'$ in Eqs.~\eqref{Z'equ} and~\eqref{Z'Ham} does %%@
not depend on time we get the solution to Eq.~\eqref{Z'equ} as
\begin{align}\label{Z'sol}
  \mathbf{Z}'(t) = &\exp{(-\mathrm{i}H't)}\mathbf{Z}'(0)
  \notag
  \\
  = & 
  \left(
    \cos\varOmega t-\mathrm{i}(\bm{\sigma}\mathbf{n})\sin\varOmega t
  \right)
  \mathbf{Z}'(0),
\end{align}
where $\mathbf{n}=(\mu\mathcal{B},0,\omega/2)/\varOmega$ is the unit vector, %%@
$\varOmega=\sqrt{(\mu\mathcal{B})^2+(\omega/2)^2}$ and $\bm{\sigma}$ are the Pauli %%@
matrices. Using Eqs.~\eqref{ZZ'matr} and~\eqref{Z'sol} we arrive to the expressions %%@
for the components of $\mathbf{Z}$ written in terms of the initial condition %%@
$\mathbf{Z}(0)$:
\begin{align}\label{Zsol}
  Z_1(t)= &
  \left(
    \cos\varOmega t-\mathrm{i}\frac{\omega}{2\varOmega}\sin\varOmega t
  \right)e^{\mathrm{i}\omega t/2}Z_1(0)
  \notag
  \\
  & -
  \mathrm{i}\frac{\mu\mathcal{B}}{\varOmega}\sin(\varOmega t)
  e^{\mathrm{i}\omega t/2}Z_2(0),
  \notag
  \\
  Z_2(t)= &
  \left(
    \cos\varOmega t+\mathrm{i}\frac{\omega}{2\varOmega}\sin\varOmega t
  \right)e^{-\mathrm{i}\omega t/2}Z_2(0)
  \notag
  \\
  & -
  \mathrm{i}\frac{\mu\mathcal{B}}{\varOmega}\sin(\varOmega t)
  e^{-\mathrm{i}\omega t/2}Z_1(0).
\end{align}
Then we identify the spinor $\mathbf{Z}$ with different functions $a_a^{(\zeta)}$ %%@
and $b_a^{(\zeta)}$, the parameters $\mathcal{B}$ and $\omega$ -- with the strength %%@
of the magnetic field and energy differences respectively (of course, with the %%@
proper signs). With help of Eqs.~\eqref{aEqsff} (and analogous equations for the %%@
functions $b_a^{(\zeta)}$) one gets the list of four cases:
\begin{itemize}
  \item For $\mathbf{Z}^\mathrm{T}=(a_1^{+{}},a_2^{+{}})$, $\mathcal{B}=-B$,
  $\omega=\omega_{+{}}$ and $\varOmega=\Omega_{+{}}$;
  \item For $\mathbf{Z}^\mathrm{T}=(a_1^{-{}},a_2^{-{}})$, $\mathcal{B}=B$,
  $\omega=\omega_{-{}}$ and $\varOmega=\Omega_{-{}}$;
  \item For $\mathbf{Z}^\mathrm{T}=(b_1^{+{}},b_2^{+{}})$, $\mathcal{B}=B$,
  $\omega=-\omega_{+{}}$ and $\varOmega=\Omega_{+{}}$;
  \item For $\mathbf{Z}^\mathrm{T}=(b_1^{-{}},b_2^{-{}})$, $\mathcal{B}=-B$,
  $\omega=-\omega_{-{}}$ and $\varOmega=\Omega_{-{}}$.
\end{itemize}
Using these formulae together with Eqs.~\eqref{Zsol} we readily arrive to %%@
Eqs.~\eqref{asol}-\eqref{Omegaomega}. Note that the dynamics of the %%@
system~\eqref{Zequ} and~\eqref{ZHam} is analogous to the quantum mechanical %%@
description of neutrino spin-flavor oscillation in a twisting magnetic field %%@
studied in Refs.~\cite{twisting}.

\section{Schr\"odinger description of spin-flavor oscillations\label{qmd}}

Let us study the Schr\"odinger evolution equation for the two mass eigenstates %%@
neutrinos with magnetic moments in an external transversal magnetic field,
\begin{align}\label{Scheq}
  \mathrm{i}\dot{\Psi}= & H\Psi,
  \notag
  \\
  H= &
  \begin{pmatrix}
    \mathcal{E}_1 & 0 & -\mu_1 B & -\mu B \\
	0 & \mathcal{E}_2 & -\mu B & -\mu_2 B \\
	-\mu_1 B & -\mu B & \mathcal{E}_1 & 0 \\
    -\mu B & -\mu_2 B & 0 & \mathcal{E}_2 \
  \end{pmatrix},
\end{align}
where mass eigenstates energies $\mathcal{E}_{1,2}$ are given in %%@
Eq.~\eqref{Ekappa}.  
The wave function of neutrinos should be presented in the following %%@
form~\cite{GriSch93}: $\Psi^\mathrm{T}=(\psi_1^\mathrm{L}, \psi_2^\mathrm{L}, %%@
\psi_1^\mathrm{R}, \psi_2^\mathrm{R})$, where $\psi_{1,2}^\mathrm{L,R}$ are %%@
one-component objects. The initial condition consistent with Eq.~\eqref{inicondpsi} %%@
is
\begin{equation}\label{inicondqm}
  \Psi^\mathrm{T}(0)=(\sin\theta,\cos\theta,0,0).
\end{equation}

To study the quantum mechanical evolution of the system we look for a solution of %%@
Eq.~\eqref{Scheq} of the form $\Psi \sim e^{-\mathrm{i}\lambda t}$. Solving the %%@
corresponding secular equation we get the eigenvalues of the Hamiltonian $H$ in the %%@
form,
\begin{equation}\label{lambda}
  \lambda_{1,2}^{+{}}=\Sigma_{+{}} \pm \Omega_{+{}},
  \quad
  \lambda_{1,2}^{-{}}=\Sigma_{-{}} \pm \Omega_{-{}},
\end{equation}
where $\Sigma_{\pm{}}=(E_1^{\pm{}}+E_2^{\pm{}})/2$ and $\Omega_{\pm{}}$ are given %%@
in Eq.~\eqref{Omegaomega}. The energy levels in an external magnetic field %%@
$E_a^{\pm{}}$ are presented in Eq.~\eqref{urenergy}. The general solution to %%@
Eq.~\eqref{Scheq} is thus takes the form
\begin{align}\label{gensolScheq}
  \Psi(t)= & 
  (\alpha_1 u_1 e^{-\mathrm{i}\Omega_{+{}}t}+ 
  \alpha_2 u_2 e^{\mathrm{i}\Omega_{+{}}t})e^{-\mathrm{i}\Sigma_{+{}}t}
  \notag
  \\
  & +
  (\beta_1 v_1 e^{-\mathrm{i}\Omega_{-{}}t}+ 
  \beta_2 v_2 e^{\mathrm{i}\Omega_{-{}}t})e^{-\mathrm{i}\Sigma_{-{}}t},
\end{align}
where the basis spinors are given as
\begin{align}\label{spinorsScheq}
  u_{1,2}= &
  \frac{1}{2\sqrt{\Omega_{+{}}}}
  \begin{pmatrix}
    R_{\pm{}} \\
	\mp \mu B/R_{\pm{}} \\
	R_{\pm{}} \\
    \mp \mu B/R_{\pm{}} \
  \end{pmatrix},
  \notag
  \\
  v_{1,2}= &
  \frac{1}{2\sqrt{\Omega_{-{}}}}
  \begin{pmatrix}
    \mp \mu B/S_{\mp{}} \\
	-S_{\mp{}} \\
	\pm \mu B/S_{\mp{}} \\
    S_{\mp{}} \
  \end{pmatrix},
\end{align}
where $R_{\pm{}}=\sqrt{\Omega_{+{}} \pm \omega_{+{}}/2}$ and %%@
$S_{\pm{}}=\sqrt{\Omega_{-{}} \pm \omega_{-{}}/2}$. The spinors $u_{1,2}$ and %%@
$v_{1,2}$ in Eq.~\eqref{spinorsScheq} are the eigenvectors of the Hamiltonian $H$ %%@
corresponding to the eigenvalues $\lambda_{1,2}^{+{}}$ and $\lambda_{1,2}^{-{}}$, %%@
respectively.

The coefficients $\alpha_{1,2}$ and $\beta_{1,2}$ should be chosen so that the %%@
initial condition in Eq.~\eqref{inicondqm} is satisfied. We choose them as
\begin{align}
  \alpha_{1,2} & =\frac{\sqrt{\Omega_{+{}}}}{2R_{\pm{}}}
  \left\{
    \sin\theta
    \left(
      1\pm\frac{\omega_{+{}}}{2\Omega_{+{}}}
    \right)
	\mp\frac{\mu B}{\Omega_{+{}}}
	\cos\theta
  \right\},
  \\
  \notag
  \beta_{1,2} & =\mp\frac{S_{\mp{}}\sqrt{\Omega_{-{}}}}{2\mu B}
  \left\{
    \sin\theta
    \left(
      1\pm\frac{\omega_{-{}}}{2\Omega_{-{}}}
    \right)
	\pm\frac{\mu B}{\Omega_{-{}}}
	\cos\theta
  \right\}.
\end{align}
Then, using Eq.~\eqref{gensolScheq}, we arrive to the following helicity components %%@
of the wave function $\Psi$:
\begin{align}\label{psiqm}
  \psi_1^\mathrm{L}(t) = &
  \frac{1}{2}
  \Big\{
    (e^{-\mathrm{i}E_1^{+{}}t}F^{+{}}+
    e^{-\mathrm{i}E_1^{-{}}t}F^{-{}})\sin\theta
	\notag
	\\
	& +
    (e^{-\mathrm{i}E_1^{+{}}t}G^{+{}}+
    e^{-\mathrm{i}E_1^{-{}}t}G^{-{}})\cos\theta
  \Big\},
  \notag
  \\
  \psi_2^\mathrm{L}(t) = &
  \frac{1}{2}
  \Big\{  
    (e^{-\mathrm{i}E_2^{+{}}t}F^{+{}*{}}+
    e^{-\mathrm{i}E_2^{-{}}t}F^{-{}*{}})\cos\theta
	\notag
	\\
	& -
    (e^{-\mathrm{i}E_2^{+{}}t}G^{+{}*{}}+
    e^{-\mathrm{i}E_2^{-{}}t}G^{-{}*{}})\sin\theta
  \Big\},
  \notag
  \\
  \psi_1^\mathrm{R}(t) = &
  \frac{1}{2}
  \Big\{
    (e^{-\mathrm{i}E_1^{+{}}t}F^{+{}}-
    e^{-\mathrm{i}E_1^{-{}}t}F^{-{}})\sin\theta
	\notag
	\\
	& +
    (e^{-\mathrm{i}E_1^{+{}}t}G^{+{}}-
    e^{-\mathrm{i}E_1^{-{}}t}G^{-{}})\cos\theta
  \Big\},
  \notag
  \\
  \psi_2^\mathrm{R}(t) = &
  \frac{1}{2}
  \Big\{  
    (e^{-\mathrm{i}E_2^{+{}}t}F^{+{}*{}}-
    e^{-\mathrm{i}E_2^{-{}}t}F^{-{}*{}})\cos\theta
	\notag
	\\
	& -
    (e^{-\mathrm{i}E_2^{+{}}t}G^{+{}*{}}-
    e^{-\mathrm{i}E_2^{-{}}t}G^{-{}*{}})\sin\theta
  \Big\},	
\end{align}
where the functions $F^{\pm{}}$ and $G^{\pm{}}$ are introduced in Eq.~\eqref{fg}.

Let us compare these wave functions with the results of our approach, presented in %%@
the main text. Initial conditions for the four-component mass eigenstates wave %%@
functions are $\psi_1(x,0)=\sin\theta e^{\mathrm{i}kx}\xi_0$ and %%@
$\psi_2(x,0)=\cos\theta e^{\mathrm{i}kx}\xi_0$ [see Eq.~\eqref{inicondpsi}], where %%@
the spinor $\xi_0$ was introduced in Sec.~\ref{PT}. It is easy to check that
\begin{align*}%\label{norhnu}
  \psi_{1,2}^\mathrm{L}(x,0)= &
  \frac{1}{2}(1-\Sigma_1)\psi_{1,2}(x,0)=\psi_{1,2}(x,0),
  \\
  \psi_{1,2}^\mathrm{R}(x,0)= &
  \frac{1}{2}(1+\Sigma_1)\psi_{1,2}(x,0)=0.
\end{align*}
These expressions imply that no right-polarized neutrinos exist initially.

With help of the following identity:
\begin{equation*}
  \frac{1}{2}(1-\Sigma_1)
  \left(
    u^{\pm{}} \otimes u^{\pm\dag} 
  \right)
  \xi_0=\frac{1}{2}\xi_0,
\end{equation*}
which can be derived from Eq.~\eqref{urspinors}, as well as using %%@
Eqs.~\eqref{xi0k0} and~\eqref{psisolff}, we obtain the helicity eigenstates of the %%@
four-component wave functions $\psi_a$, $a=1,2$,
\begin{align}\label{psiqft}
  \psi_1^\mathrm{L}(t) = &
  \frac{1}{2}
  \Big\{
    (e^{-\mathrm{i}E_1^{+{}}t}F^{+{}}+
    e^{-\mathrm{i}E_1^{-{}}t}F^{-{}})\sin\theta
	\\
	& +
    (e^{-\mathrm{i}E_1^{+{}}t}G^{+{}}+
    e^{-\mathrm{i}E_1^{-{}}t}G^{-{}})\cos\theta
  \Big\}
  e^{\mathrm{i}kx}\xi_0,
  \notag  
  \\
  \psi_2^\mathrm{L}(t) = &
  \frac{1}{2}
  \Big\{  
    (e^{-\mathrm{i}E_2^{+{}}t}F^{+{}*{}}+
    e^{-\mathrm{i}E_2^{-{}}t}F^{-{}*{}})\cos\theta
	\notag
	\\
	& -
    (e^{-\mathrm{i}E_2^{+{}}t}G^{+{}*{}}+
    e^{-\mathrm{i}E_2^{-{}}t}G^{-{}*{}})\sin\theta
  \Big\}
  e^{\mathrm{i}kx}\xi_0,
  \notag
  \\
  \psi_1^\mathrm{R}(t) = &
  \frac{1}{2}
  \Big\{
    (e^{-\mathrm{i}E_1^{+{}}t}F^{+{}}-
    e^{-\mathrm{i}E_1^{-{}}t}F^{-{}})\sin\theta
	\notag
	\\
	& +
    (e^{-\mathrm{i}E_1^{+{}}t}G^{+{}}-
    e^{-\mathrm{i}E_1^{-{}}t}G^{-{}})\cos\theta
  \Big\}
  e^{\mathrm{i}kx}\kappa_0,
  \notag
  \\
  \psi_2^\mathrm{R}(t) = &
  \frac{1}{2}
  \Big\{  
    (e^{-\mathrm{i}E_2^{+{}}t}F^{+{}*{}}-
    e^{-\mathrm{i}E_2^{-{}}t}F^{-{}*{}})\cos\theta
	\notag
	\\
	& -
    (e^{-\mathrm{i}E_2^{+{}}t}G^{+{}*{}}-
    e^{-\mathrm{i}E_2^{-{}}t}G^{-{}*{}})\sin\theta
  \Big\}
  e^{\mathrm{i}kx}\kappa_0,
  \notag
\end{align}
where the spinor $\kappa_0$ is given in Eq.~\eqref{Ekappa}. The comparison of %%@
Eq.~\eqref{psiqft} with Eq.~\eqref{psiqm} shows that Schr\"odinger approach gives %%@
results analogous to the results obtained in our approach based on Pauli-Dirac %%@
equation.

\end{document}